\documentclass[reprint,superscriptaddress,secnumarabic,amssymb, nobibnotes, aps, prl]{revtex4-2}
\usepackage{setspace}
\usepackage{gensymb}
\usepackage{siunitx}
\usepackage{mathrsfs}
\usepackage{amsmath}
\usepackage[export]{adjustbox}
\usepackage{caption}
\usepackage{tikz}
\usepackage{xr}

\makeatletter
\newcommand*{\addFileDependency}[1]{
  \typeout{(#1)}
  \@addtofilelist{#1}
  \IfFileExists{#1}{}{\typeout{No file #1.}}
}
\makeatother

\newcommand*{\myexternaldocument}[1]{%
    \externaldocument{#1}%
    \addFileDependency{#1.tex}%
    \addFileDependency{#1.aux}%
}

\myexternaldocument{SupplementalMaterials}

\usepackage[labelformat=simple]{subcaption}
\usepackage{braket}
\usepackage{graphicx}
\graphicspath{ {./Figures/} }

\newcommand{\RNum}[1]{\uppercase\expandafter{\romannumeral #1\relax}}

\begin{document}

\title{Experimental electronic structure of the electrically switchable antiferromagnet CuMnAs}%

\author{A. Garrison Linn}
\email[Corresponding Author: ]{Garrison.Linn@colorado.edu}
\affiliation{Department of Physics, University of Colorado at Boulder, Boulder, CO 80309, USA}
\author{Peipei Hao}
\affiliation{Department of Physics, University of Colorado at Boulder, Boulder, CO 80309, USA}
\author{Kyle N. Gordon}
\affiliation{Department of Physics, University of Colorado at Boulder, Boulder, CO 80309, USA}
\author{Dushyant Narayan}
\affiliation{Department of Physics, University of Colorado at Boulder, Boulder, CO 80309, USA}
\author{Bryan S. Berggren}
\affiliation{Department of Physics, University of Colorado at Boulder, Boulder, CO 80309, USA}
\author{Nathaniel Speiser}
\affiliation{Department of Physics, University of Colorado at Boulder, Boulder, CO 80309, USA}
\author{Sonka Reimers}
\affiliation{School of Physics and Astronomy, University of Nottingham, University Park, Nottingham NG7 2RD, United Kingdom}
\affiliation{Diamond Light Source, Harwell Science and Innovation Campus, Didcot OX11 0DE, United Kingdom}
\author{Richard P. Campion}
\affiliation{School of Physics and Astronomy, University of Nottingham, University Park, Nottingham NG7 2RD, United Kingdom}
\author{V\'{i}t Nov\'{a}k}
\affiliation{Institute of Physics, Academy of Sciences of the Czech Republic, Cukrovarnicka 10, 162 00 Praha 6, Czech Republic}
\author{Sarnjeet S. Dhesi}
\affiliation{Diamond Light Source, Harwell Science and Innovation Campus, Didcot OX11 0DE, United Kingdom}
\author{Timur Kim}
\affiliation{Diamond Light Source, Harwell Science and Innovation Campus, Didcot OX11 0DE, United Kingdom}
\author{Cephise Cacho}
\affiliation{Diamond Light Source, Harwell Science and Innovation Campus, Didcot OX11 0DE, United Kingdom}
\author{Libor \v{S}mejkal}
\affiliation{Institute of Physics, Academy of Sciences of the Czech Republic, Cukrovarnicka 10, 162 00 Praha 6, Czech Republic}
\affiliation{Institut f\"ur Physik, Johannes Gutenberg-Universit\"{a}t of Mainz, 55128 Mainz, Deutschland}
\author{Tom\'{a}\v{s} Jungwirth}
\affiliation{School of Physics and Astronomy, University of Nottingham, University Park, Nottingham NG7 2RD, United Kingdom}
\affiliation{Institute of Physics, Academy of Sciences of the Czech Republic, Cukrovarnicka 10, 162 00 Praha 6, Czech Republic}
\author{Jonathan D. Denlinger}
\affiliation{Advanced Light Source, Lawrence Berkeley National Laboratory, Berkeley, California 94720, USA}
\author{Peter Wadley}
\affiliation{School of Physics and Astronomy, University of Nottingham, University Park, Nottingham NG7 2RD, United Kingdom}
\author{Dan Dessau}
\affiliation{Department of Physics, University of Colorado at Boulder, Boulder, CO 80309, USA}
\affiliation{Center for Experiments on Quantum Materials, Boulder, CO 80309, USA}
\date{\today}%

\begin{abstract}
Tetragonal CuMnAs is a room temperature antiferromagnet with an electrically reorientable N\'eel vector and a Dirac semimetal candidate. Direct measurements of the electronic structure of single-crystalline thin films of tetragonal CuMnAs using angle-resolved photoemission spectroscopy (ARPES) are reported, including Fermi surfaces (FS) and energy-wavevector dispersions. After correcting for a chemical potential shift of $\approx-390$~meV (hole doping), there is excellent agreement of FS, orbital character of bands, and Fermi velocities between the experiment and density functional theory calculations. Additionally, 2x1 surface reconstructions are found in the low energy electron diffraction (LEED) and ARPES. This work underscores the need to control the chemical potential in tetragonal CuMnAs to enable the exploration and exploitation of the Dirac fermions with tunable masses, which are predicted to be above the chemical potential in the present samples.
\end{abstract}

\maketitle

\begin{figure}[t]
    \includegraphics{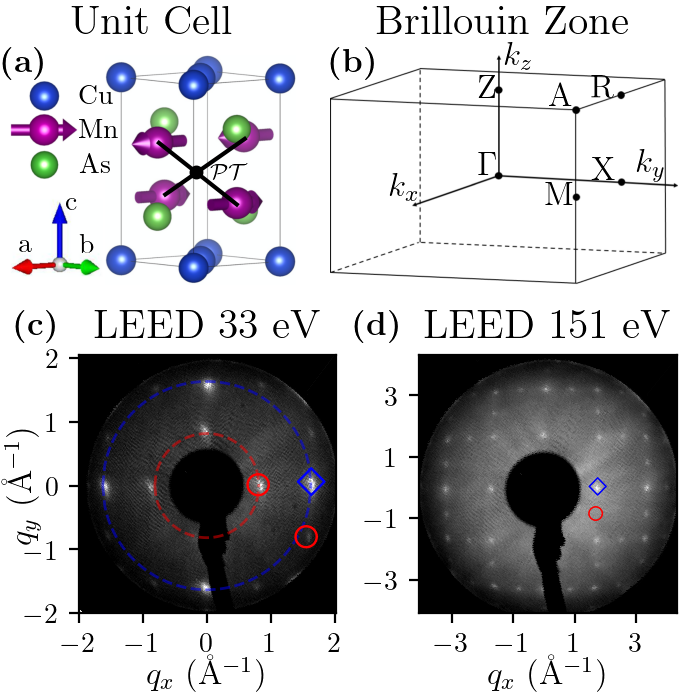}
	\caption{CuMnAs unit cell, Brillouin zone, and LEED: (a) Real space unit cell depicting the magnetic moments on the Mn atoms with arrows. (b) Tetragonal Brillouin zone, labeling high symmetry points. (c\&d) LEED patterns from a decapped sample taken with $33$~eV and $151$~eV electrons, respectively. The blue diamond encloses an expected peak from tetragonal CuMnAs, while the red circles enclose extra peaks from 2x1 and 1x2 surface reconstruction domains. (c) The blue (red) dashed circle is drawn with a radius of $\frac{2\pi}{a}$ ($\frac{2\pi}{2a}$) where a is the in-plane lattice constant of CuMnAs.}\label{Fig1}
\end{figure}

CuMnAs has emerged as an exciting material, both for spintronic applications~\cite{WadleyElectricalSwitching,STDevice} and the study of anti-ferromagnetic (AFM) Dirac materials~\cite{smejkal_topological_2018,DiracAFM,SmejkalTheory}. CuMnAs has a combined inversion and time reversal symmetry, $\mathcal{PT}$, that connects two oppositely oriented magnetic Mn sublattices with a N\'eel temperature ${\sim}480$~K [see Fig.~\ref{Fig1}(a)]~\cite{wadley_tetragonal_2013,RoomTemp}. This $\mathcal{PT}$ symmetry, highlighted in Fig.~\ref{Fig1}(a), in conjunction with the relativistic spin-orbit coupling allows a current to induce a non-equilibrium spin-polarization that is staggered and commensurate with the equilibrium AFM order. The non-equilibrium and equilibrium AFM moments couple to each other resulting in a spin orbit torque that reorients the N\'eel vector. Therefore, current driven through CuMnAs can efficiently reorient the N\'eel vector. This was theoretically predicted~\cite{SOTPrediction,SmejkalTheory} and experimentally confirmed in CuMnAs and in another $\mathcal{PT}$ symmetric antiferromagnet, Mn$_2$Au~\cite{XMLD_Switch, WadleyElectricalSwitching, THzSwitchingSpeeds,Bodnar}. For example, photoemission electron microscopy with x-ray magnetic linear dichroism providing contrast was used to directly image the N\'eel vector reorientation after passing currents through tetragonal CuMnAs~\cite{XMLD_Switch}. 

Control over the N\'eel vector allows manipulation of some electronic properties of CuMnAs. First, the presence of magnetic order gives rise to an anisotropic magnetoresistance (AMR). Therefore, the resistivity along the $a$ lattice direction can be modulated by orienting the N\'eel vector to be perpendicular or parallel to $a$, for example. Wadley et al. made such a device out of a thin film of tetragonal CuMnAs and used the AMR signal to demonstrate the electrical switching of the N\'eel vector~\cite{WadleyElectricalSwitching}. The electrical switching was even shown to be scalable to THz speeds~\cite{THzSwitchingSpeeds}. Second, the $\mathcal{PT}$ symmetry provides doubly degenerate bands, which allows for the existence of antiferromagnetic Dirac fermions close to or at the Fermi level~\cite{SmejkalTheory}. Additional off-centered or nonsymmorphic crystallographic symmetries can protect the massless Dirac fermions. The presence or absence of these symmetries can be controlled by reorienting the N\'eel vector resulting in tunable masses of the Dirac fermions. The opening and closing of the Dirac mass gap results in enhancement of the AMR\cite{SmejkalTheory, Bodnar}. 

Both the tetragonal and orthorhombic phases of CuMnAs have been studied theoretically and experimentally. All of the above interesting properties are believed to be present in both structural phases. Additionally, the orthorhombic phase is proposed to host a new topological metal-insulator transition, due to the predicted presence of a bulk Dirac point at $E_F$ and lack of other bands crossing $E_F$~\cite{SmejkalTheory}. 

Density Functional Theory (DFT) has been critical to the development of the above theoretical predictions, but it has only been experimentally tested to a limited extent in tetragonal CuMnAs: the AC permittivity (determined from ellipsometry) and UV photoemission spectroscopy were studied~\cite{PESPapere}; neutron diffraction and x-ray magnetic linear dichroism were used to study the magnetic ordering~\cite{wadley_antiferromagnetic_2015}; and, more recently, experiments using scanning tunneling microscopy elucidated the surface termination of thin films of tetragonal CuMnAs, discovering the existence of As step edges which may host surface reconstructions~\cite{STM}. These experiments were compared to DFT predictions; however, there exist no direct comparisons to the band structure calculated from DFT.

Therefore, to directly probe the band structure of CuMnAs, high resolution ARPES measurements were made at the MERLIN ARPES endstation of beamline 4.0.3 at the Advanced Light Source and at the HR-ARPES branch of beamline i05 at the Diamond Light Source. The base pressure was $\lesssim5\times10^{-11}$ Torr, with temperatures below 50K. Samples were 45 nm thick films of single-crystalline tetragonal CuMnAs with the (001) face exposed and grown on GaP(001). The films were capped with 30 nm of As to protect the surfaces from contamination from the ambient environment. Decapping was performed in an environment with pressures $\lesssim10^{-8}$ Torr, reaching a max temperature reading of $340$~$^{\circ}$C on a pyrometer, emissivity $=0.1$. 

After decapping, LEED was performed \emph{in situ}, revealing a well ordered surface with reconstructions [see Fig.~\ref{Fig1}(c\&d)]. The (001) surface of tetragonal CuMnAs should produce a square reciprocal lattice with edge lengths of $2\pi/a$, where $a$ is determined from XRD to be $3.85$~\si{\angstrom}. This is seen in the LEED patterns; however, there are additional spots at the midpoints along the edges of the squares, suggesting the presence of 2x1 and 1x2 surface reconstruction domains. The reconstruction is confirmed by observing that the first order Bragg peaks occur at a radius of $2\pi/2a$, where without a reconstruction they would occur at $2\pi/a$ [see Fig.~\ref{Fig1}(c)]. Despite the surface reconstruction, the LEED pattern is sharp, indicating a successful decap. Samples were subsequently transferred into the ARPES chamber, maintaining an ultra-high vacuum environment from decap to ARPES data acquisition.

We show the crystallographic and magnetic structure of tetragonal phase of CuMnAs in Fig.~\ref{Fig1}(a). The nonmagnetic space group is nonsymmoprhic (P4/nmm). Since the magnetic Mn atoms are light, the effects of spin-orbit coupling on the band-structure are highly perturbative and smaller than our resolution can detect, unlike in strongly relativistic $\mathcal{PT}$ antiferromagnet Mn$_2$Au~\cite{Bodnar_Parity}. Therefore, in the antiferromagnetic state the nonrelativsitic spin group (P14/$^2$n$^2$m$^2$m) describes the main energy scales of our measured band structure, including Fermi surfaces~\cite{Libor_PRX}. However, within ${\sim}10$~meV of a Dirac point, the fermion masses are required to accurately describe the electronic dispersion, and to calculate the Dirac fermion masses, spin-orbit coupling must be included and the relativistic magnetic symmetry group (Pm'mn) with generators $\{C_{2x}|\frac{1}{2} 0 0\}$, $\{M_{y}|0 \frac{1}{2} 0\}$, and $\mathcal{PT}$ must be employed. 

To model the electronic structure of tetragonal CuMnAs, DFT was performed, using the Generalized Gradient Approximations (GGA) with the Coulomb interaction $U$ applied to the Mn 3d orbitals within the Dudarev approximation~\cite{VASP} [see Sec.~I of Supplemental Materials (SM) for a complete description of our DFT]. For the reasons stated above, spin-orbit coupling was turned off for all DFT shown, except Fig.~\ref{SymmetryCuts}(g,h,\&i) and Fig.~S2 in SM. The best quantitative agreement of Fermi velocities was found for $U=2.25$~eV, while maintaining excellent overall qualitative agreement. It was necessary to apply $\approx390$~meV of rigid hole doping to the chemical potential from theory, i.e. the experimental Fermi energy was found to be equal to the theoretical Fermi energy $-390$~meV. Possible origins of the energy shift will be discussed. Therefore, unless explicitly discussed, all of the DFT shown in this manuscript has been plotted after applying the above mentioned rigid chemical potential shift.

\begin{figure*}
    \includegraphics{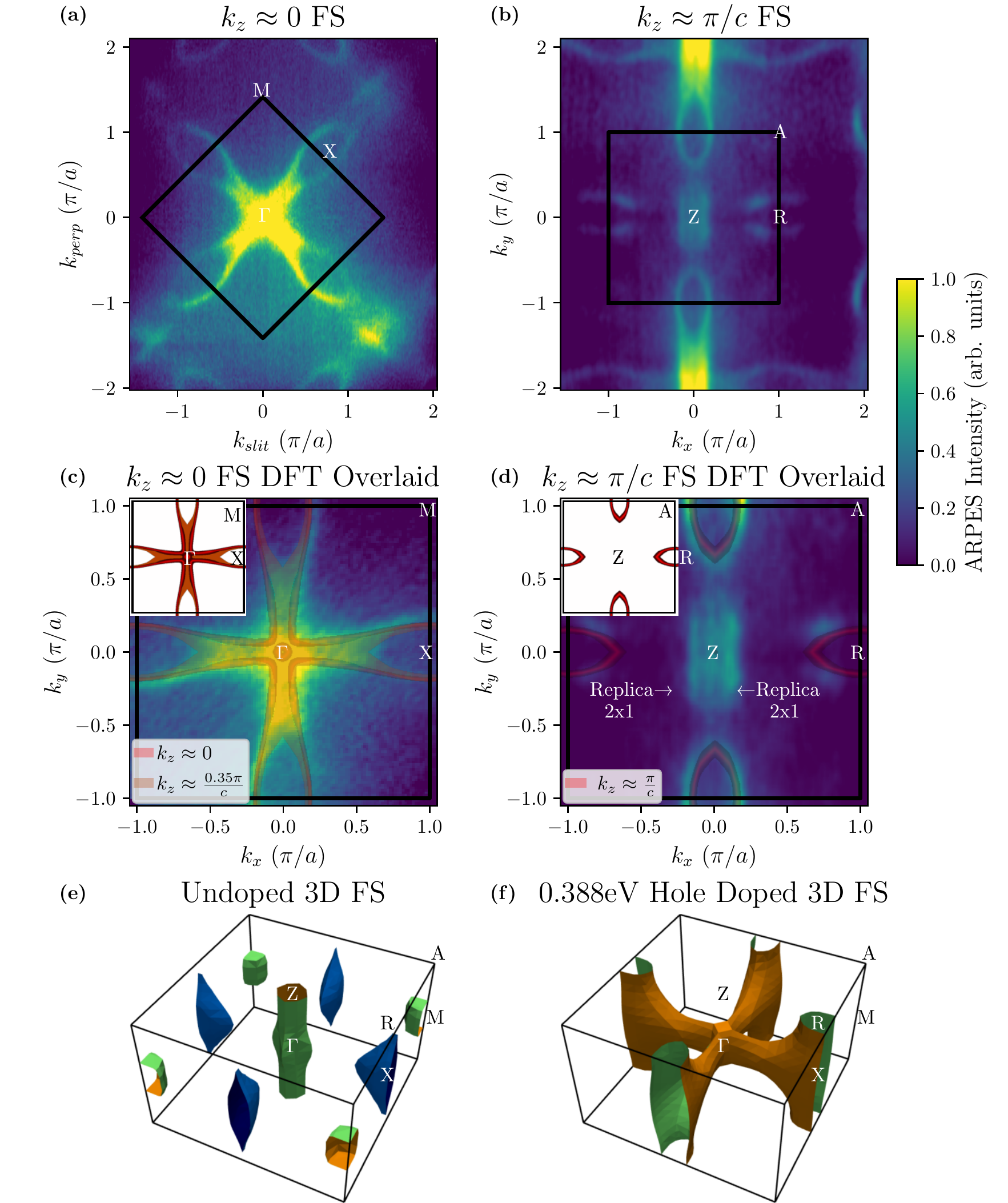}
    \caption{Fermi surface: Unsymmetrized Fermi surface data taken with $85$~eV (a\&c) and $100$~eV (b\&d), linear vertically polarized photons. (a-d) The same data is shown twice. The black box depicts the Brillouin zone boundary. The data is from the CuMnAs01 sample. (c\&d) The DFT is overlaid with a transparency. The insets show only the DFT. (e\&f) 3D Fermi surfaces from the DFT plotted with PyProcar~\cite{PyProcar}: (e) undoped and (f) doped.}\label{FS}
\end{figure*}

\begin{figure*}[h]
    \includegraphics{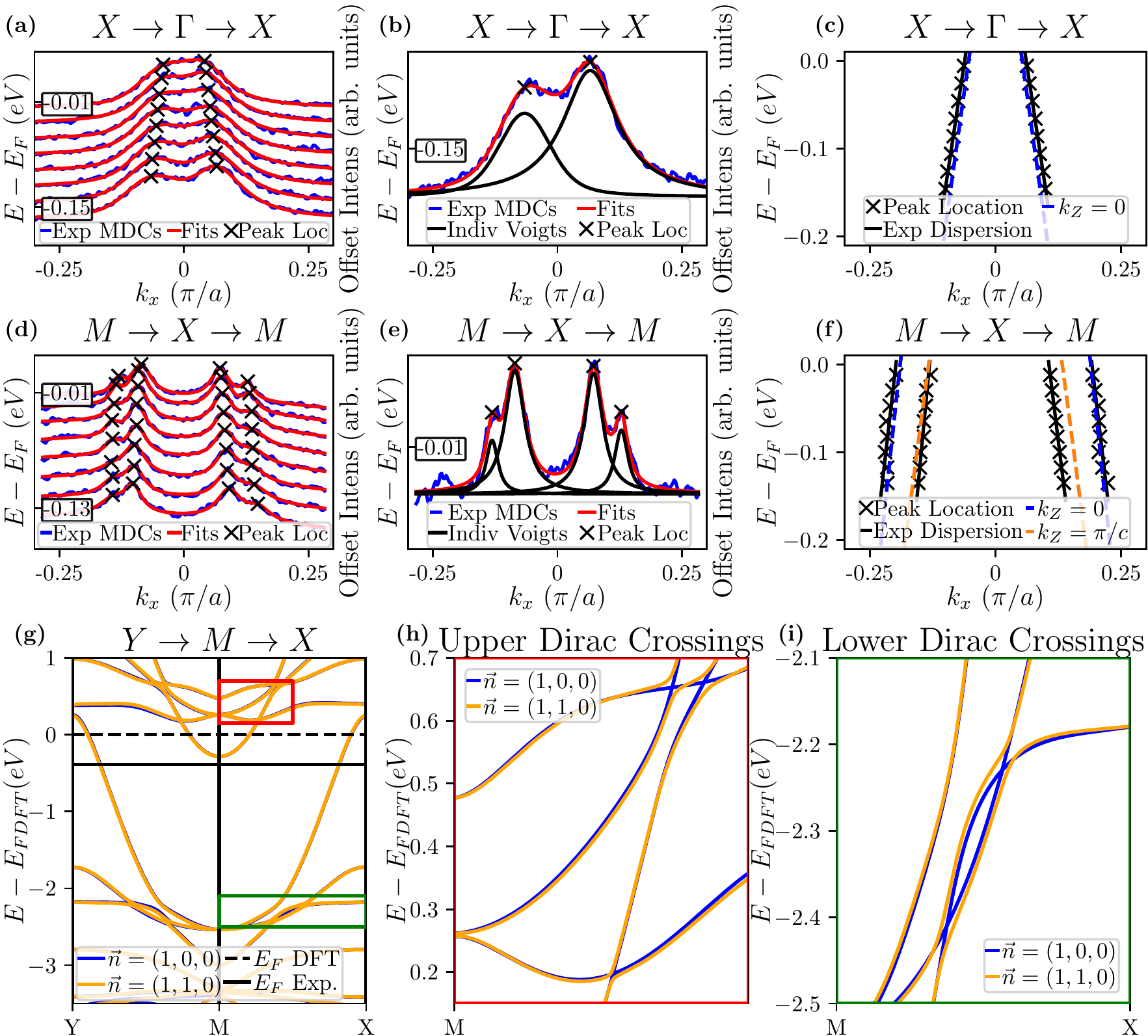}
    \caption{Symmetry Cuts: (a\&d) Waterfall plots of MDCs with fits overlaid. (b\&e) Single MDC from (a\&d), highlighting individual components of fits. (c\&f) Comparison of the dispersion from the DFT and experiment. (a-f) The $X\to\Gamma\to X$ and $M\to X\to M$ symmetry cuts––(a-c) and (d-f), respectively––were acquired with $85$~eV, LV polarized photons. The $X\to\Gamma\to X$ data comes from the CuMnAs01 sample and the $M\to X\to M$ data from the CuMnAs02 sample. (g) Shows the $Y\to M\to X$ cut Near $E_F$ from DFT with spin-orbit coupling included. (h\&i) zoom in on the dispersion in the red and green boxes shown in (g), respectively, highlighting nearby Dirac fermions with electrically tunable mass gaps. }\label{SymmetryCuts}
\end{figure*}

As a first step to understanding the electronic structure of tetragonal CuMnAs, the in-plane Fermi surface was measured at $k_z\approx 0$ [see Fig.~\ref{FS}(a)] and $k_z\approx\pi/c$ [see Fig.~\ref{FS}(b)], which may be selected by tuning the photon energy (see Sec.~IV of SM). In this case, the zone center data were taken with $85$~eV photons, whereas the zone edge data were taken at $100$~eV. In both cases, the Fermi surface shows strong agreement with the DFT—they both show a propeller-shaped Fermi surface in the $k_z  \approx 0$ plane and ellipses at the $R$ points in the $k_z  \approx \pi/c$ plane, with the experiment and theory displaying a very similar shape and size of these features [see Fig.~\ref{FS}(c) and Fig.~\ref{FS}(d)]. There are, however, three subtle effects that might naively appear as qualitative discrepancies between the experiment and theory: First, unlike the DFT, the data shown in Fig.~\ref{FS}(b\&d) does not have $C_4$ symmetry. While turning on spin-orbit coupling in the DFT would break this symmetry, the affect is too small to explain the data or even detect (see Fig.~S2 of SM for Fermi surfaces with spin-orbit coupling included). Instead, it is primarily the transition matrix element that is known to modulate the ARPES spectral intensity that is responsible for breaking the $C_4$ symmetry in the data. In fact, a detailed analysis of this matrix element effect reveals the orbital character of the Fermi surface, which is in good agreement with that of the DFT (see Sec.~V of SM). Second, due to the inherent surface sensitivity of ARPES and lack of translational symmetry perpendicular to the surface, ARPES spectra from any photon energy can contain contributions from multiple $k_z$ values. This explains the closed ellipses near the $X$ points in Fig.~\ref{FS}(a) and some of the weight at $k_x=0$ near the $Z$ point in Fig.~\ref{FS}(b). This point is visually illustrated in Fig.~\ref{FS}(c) by overlaying the DFT from a second $k_z$ value (orange transparency) that is $0.35\pi/c$ away from the anticipated $k_z$ value (red transparency). Third, there is extra spectral weight near the $Z$ point in Fig.~\ref{FS}(d) indicated by white text, which is identified as a replica of the vertical ellipses enclosing the zone edges at $k_x=0$. The back-folding of this ellipse onto $Z$ is the analogue of the 2x1 surface reconstructions observed in the LEED and will be analyzed quantitatively later.

To clearly illustrate that the DFT must be hole doped to be consistent with the experimental data, the undoped and doped Fermi surface from the DFT are shown in Fig.~\ref{FS}(e) and Fig.~\ref{FS}(f), respectively. First, the experimental Fermi surface does not show pockets near the $M$ points. Second, the pocket near the $\Gamma$ point clearly connects with the pocket near the $X$ point in the experimental data. These points are inconsistent with the undoped Fermi surface. However, after lowering the chemical potential of the DFT by $\approx0.390$~eV (rigid hole doping), all of the qualitative and even quantitative features of the experimental Fermi surface are consistent with the bulk DFT calculations.

With the aid of Fermi surface plots and $k_z$ dispersion (see Sec.~IV of SM~IV), $E$-$k$ dispersion for the high symmetry cuts in the $\Gamma$-$X$ plane are readily acquired using $85$~eV, LV polarized photons. To quantitatively compare the experimental data to DFT, momentum distribution cuts (MDCs) are fit to extract the low energy experimental dispersion, including Fermi velocities [see Fig.~\ref{SymmetryCuts}(a-f)]. The $X \to \Gamma \to X$ MDCs are fit with two Voigt functions, and the extracted Fermi velocity is $4.6$~eV\,\si{\angstrom}, which to within the experimental error bars is the same as the Fermi velocity from the DFT, $v_{F} = 4.8$~eV\,\si{\angstrom}. The $M \to X \to M$ symmetry cut contains the $A \to R \to A$ bands near $E_F$, due to the same $k_z$ uncertainty mentioned above. Therefore, the extracted MDCs were fit with four Voigt functions, representing four bands. The Fermi velocity of the bands corresponding to the $M \to X \to M$ cut is $6.0$~eV\,\si{\angstrom}. Again to within experimental precision, this is the same as the Fermi velocity from the DFT, $v_{F} = 6.4$~eV\,\si{\angstrom}. The error on the extracted Fermi velocities from MDC fittings scales as $\sim {v_{F}}^{2}$. So, for these large Fermi velocities, the relative error is found to be $\sim 10\%$ (see Sec.~VII of SM). To see the full ARPES images for several high symmetry cuts along with a matrix element analysis of the cuts shown, see Sec.~VI of SM. 

The strong agreement between the DFT and experiment shown throughout this paper lend credence to the predictions from the DFT beyond the ones directly verified. For example, this applies to the atomic character of the bands predicted by the DFT [see Fig.~S1 of SM]. More importantly, the DFT predicts the presence of Dirac fermions with a tunable mass gap, which are present in the DFT when spin-orbit coupling is included [see Fig.~\ref{SymmetryCuts}(g-i)]. The closest of which is just $180$~meV above the stoichiometric and defect free Fermi energy. Note that moving $E_F$ to this Dirac point should enable band topology switching. 

\begin{figure}
    \includegraphics{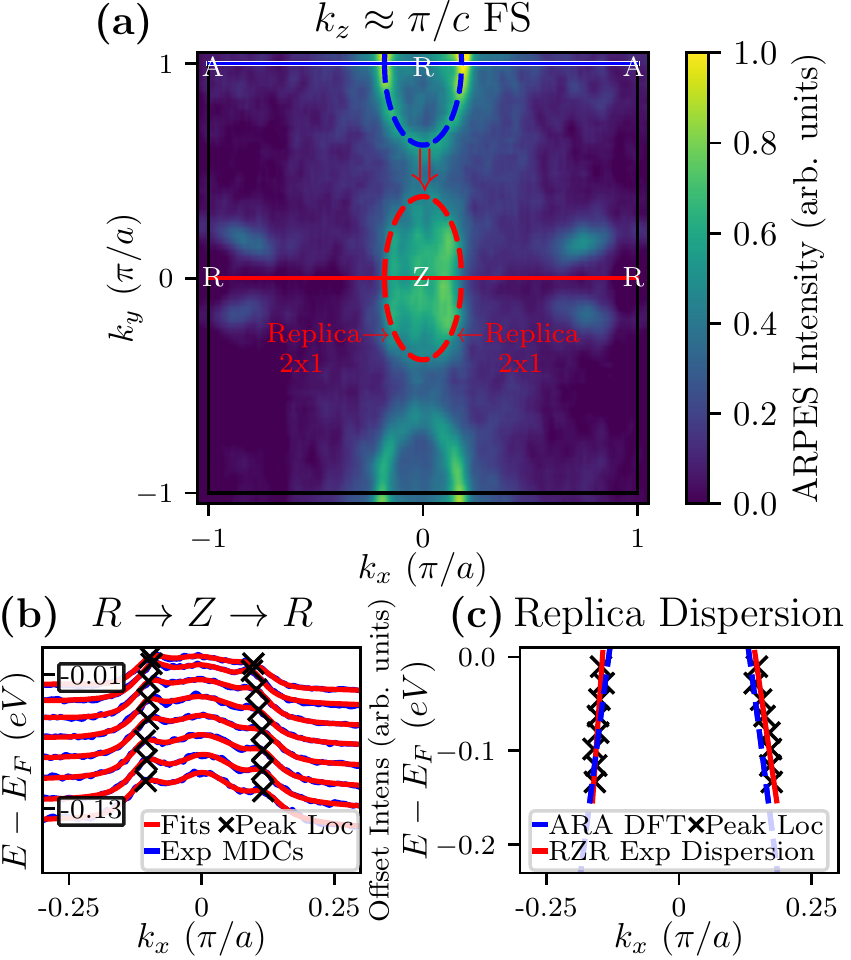}
    \caption{Replica band: (a) The same Fermi surface data as shown in Fig.~\ref{FS}(b). The blue dashed ellipse represents a primary pocket that replicates as the red dashed ellipse due to the 1x2 surface reconstructions. (b) Waterfall plot of MDCs extracted from the $R\to Z\to R$ symmetry cut [red line in (a)]. (c) Comparison of the experimental $R\to Z\to R$ dispersion and the $A\to R \to A$ dispersion from the DFT. The blue line in (a) shows the $A\to R \to A$ symmetry cut.}\label{Surface Reconstruction}
\end{figure}

Having demonstrated the strong agreement between the DFT and experiment, it is time to address two issues raised previously in detail—the replication and doping. First, to quantitatively test the hypothesis that the extra structure near the $Z$ point of the Fermi surface is the replication of the ellipses near the $R$ point [see Fig.~\ref{Surface Reconstruction}(a)], the dispersion along the $R\to Z\to R$ cut is compared to the dispersion along the $A\to R\to A$ cut. MDC fitting is used to extract the experimental dispersion along the $R\to Z\to R$ cut [see Fig.~\ref{Surface Reconstruction}(b)]. The MDCs are fit with three Voigt functions—two to capture the bands in question and one to capture a band top at $k_x=0$ that disperses across $E_F$ as a function of $k_z$. The extracted dispersion from the $R\to Z\to R$ cut is overlaid with the DFT dispersion along the $A\to R\to A$ cut, showing excellent qualitative agreement [see Fig.~\ref{Surface Reconstruction}(c)]. From MDC analysis, the Fermi velocity of the bands in the $R\to Z \to R$ cut is determined to be $5.1$~eV\,\si{\angstrom}. The Fermi velocity of the $A\to R\to A$ cut extracted from the DFT is $5.2$~eV\,\si{\angstrom}, which is within $2\%$ of the experimental value.

Second, to determine if the $\approx390$~meV chemical potential shift is reasonable, necessary concentrations of likely defects are calculated. It turns out that two effects allow this relatively large energy shift to be explained by reasonably small defect concentrations. First, CuMnAs is a semimetal. According to the DFT, $\approx390$~meV of hole doping corresponds to only $0.27$ holes per unit cell. Second, the most likely defects all bring a substantial number of holes with each defect. According to M\'aca et al., the most energetically favorable defects are Cu and Mn vacancies followed by Mn substitutions for Cu~\cite{Maca}. One reasonably suspects that the non-magnetic copper will have 10 valence electrons, i.e. will be [Ar] $3$d$^{10}$, and the magnetic Mn will be [Ar] $3$d$^{5}$ with 5 valence electrons, naively leaving As with a full valence shell. The atomically projected Density Of States (DOS) from the DFT mostly agrees with this intuition, finding 10 valence electrons on Cu and 5.4 on Mn, which would mean that each Mn for Cu substitution brings 4.6 holes. Therefore, taking into account the 222 stoichiometry of the unit cell, the $388$~meV shift would result from only $1.3\%$, $2.4\%$, and $2.9\%$ of pure Cu or Mn vacancies or Mn for Cu substitutions, respectively. Furthermore, there could be combinations of these defects, yielding very reasonable defect levels. Specifically, the defect concentrations in the few percent range give, according to Máca at al.~\cite{Maca}, DFT resistivities of CuMnAs consistent with experiment.

In-plane Fermi surfaces, $k_z$ dispersions (see Sec.~IV of SM), and symmetry cuts as $E$-$k$ dispersions for tetragonal CuMnAs are reported for the first time. After shifting the chemical potential by $\approx-390$~meV (hole doping), DFT calculations—using GGA+U with $U=2.25$~eV applied to the Mn-3d orbitals—are found to be in excellent qualitative and quantitative agreement with the experimental data. In particular, the DFT predicts accurate Fermi velocities and an orbital character for the bands that is consistent with the experimental results. Additionally, surface reconstruction and replicated bands are found.

Furthermore, the extracted value of $U$ is consistent with that of other studies. Veis et al. found that $U_{eff}=(U-J)\approx2$~eV fit their angle-integrated photoemission data best~\cite{PESPapere}. In the GGA+U scheme used for the calculations in this paper, $U$ and $J$ values are not separate, and what really enters the total energy is the ``$U-J$'' value. So, the $U=2.25$~eV term in this manuscript corresponds to the $U_{eff}$. Guyen et al. used $U=2$~eV to explain emergent edge states on their surface reconstructed CuMnAs thin films~\cite{STM}. $U<2$~eV fails to even qualitatively capture the ARPES experimental results, primarily because the bandwidth of the band in the $X\to\Gamma\to X$ is too small. The strongly constrained and appropriately normed (SCAN) functional~\cite{SCAN} increased this bandwidth compared to pure GGA, yielding a similar bandwidth to $U=1$~eV. Nevertheless, SCAN still did not produce a large enough bandwidth for the $X\to\Gamma\to X$ cut. $U=2.25$~eV within GGA was primarily chosen because it reproduces the Fermi velocity in the $X\to\Gamma\to X$ cut most accurately.

By showing that DFT accurately captures the electronic structure of tetragonal CuMnAs, more weight is given to the growing interest in CuMnAs as a candidate AFM topological Dirac material. Furthermore, the low DOS near $E_F$ provides hope that tetragonal CuMnAs can be electron doped moving $E_F$ into the proximity of the electrically switchable Dirac points above $E_F$ in the studied films. According to the DFT, the Dirac points are $570$~meV or ${\sim}0.38$ electrons per unit cell above the chemical potential in the current films or only $180$~meV/${\sim}0.12$ electrons per unit cell above the defect free chemical potential. Alternatively, orthorhombic CuMnAs was suggested to be a pristine antiferromagnetic Dirac semimetal with only Dirac fermions at the Fermi level without any trivial bands. Considering the reliability of DFT in the description of tetragonal CuMnAs presented here, studying orthorhombic CuMnAs appears as a promising direction in research of antiferromagnetic Dirac semimetals. 

A. G. Linn and K. N. Gordon contributed equally to gathering the ARPES data with support from J. D. Denlinger, P. Hao, B. S. Berggren, D. Narayan, T. Kim, C. Cacho, N. Speiser, S. Reimers, and S. Dhesi. The DFT shown in this paper was performed by P. Hao with feedback from A. G. Linn, D. Dessau, L. \v{S}mejkal, and T. Jungwirth. P. Wadley oversaw the growth and characterization of the thin films of tetragonal CuMnAs by R.P. Campion and V. Novak. This manuscript was prepared by A. G. Linn under guidance from Dan Dessau and with feedback from all authors. Dan Dessau was the principal investigator overseeing the work. 

This work was supported by DOE grant DE-FG02-03ER46066, Betty and Gordon Moore Foundation grant GBMF9458, Ministry of Education of the Czech Republic Grants LNSM-LNSpin, LM2018140, Czech Science Foundation Grant No. 19-28375X, and EU FET Open RIA Grant No. 766566. Additionally, P.Wadley acknowledges support from the Royal Society through a Royal Society University Research Fellowship. This research used resources of the Advanced Light Source, a U.S. DOE Office of Science User Facility under contract no. DE-AC02-05CH11231. We would like to thank the Diamond i05 ARPES (SI22665, SI24224-1) and ALS Merlin ARPES end station teams for allowing us time for and aiding in the ARPES data acquisition.

\bibliographystyle{unsrt}
\bibliography{./main.bbl}

\newpage

\end{document}


\title{Supplemental Materials: Experimental electronic structure of the electrically switchable antiferromagnet CuMnAs}%

\author{A. Garrison Linn}
\email[Corresponding Author: ]{Garrison.Linn@colorado.edu}
\affiliation{Department of Physics, University of Colorado at Boulder, Boulder, CO 80309, USA}
\author{Peipei Hao}
\affiliation{Department of Physics, University of Colorado at Boulder, Boulder, CO 80309, USA}
\author{Kyle N. Gordon}
\affiliation{Department of Physics, University of Colorado at Boulder, Boulder, CO 80309, USA}
\author{Dushyant Narayan}
\affiliation{Department of Physics, University of Colorado at Boulder, Boulder, CO 80309, USA}
\author{Bryan S. Berggren}
\affiliation{Department of Physics, University of Colorado at Boulder, Boulder, CO 80309, USA}
\author{Nathaniel Speiser}
\affiliation{Department of Physics, University of Colorado at Boulder, Boulder, CO 80309, USA}
\author{Sonka Reimers}
\affiliation{School of Physics and Astronomy, University of Nottingham, University Park, Nottingham NG7 2RD, United Kingdom}
\affiliation{Diamond Light Source, Harwell Science and Innovation Campus, Didcot OX11 0DE, United Kingdom}
\author{Richard P. Campion}
\affiliation{School of Physics and Astronomy, University of Nottingham, University Park, Nottingham NG7 2RD, United Kingdom}
\author{V\'{i}t Nov\'{a}k}
\affiliation{Institute of Physics, Academy of Sciences of the Czech Republic, Cukrovarnicka 10, 162 00 Praha 6, Czech Republic}
\author{Sarnjeet S. Dhesi}
\affiliation{Diamond Light Source, Harwell Science and Innovation Campus, Didcot OX11 0DE, United Kingdom}
\author{Timur Kim}
\affiliation{Diamond Light Source, Harwell Science and Innovation Campus, Didcot OX11 0DE, United Kingdom}
\author{Cephise Cacho}
\affiliation{Diamond Light Source, Harwell Science and Innovation Campus, Didcot OX11 0DE, United Kingdom}
\author{Libor \v{S}mejkal}
\affiliation{Institute of Physics, Academy of Sciences of the Czech Republic, Cukrovarnicka 10, 162 00 Praha 6, Czech Republic}
\affiliation{Institut f\"ur Physik, Johannes Gutenberg-Universit\"{a}t of Mainz, 55128 Mainz, Deutschland}
\author{Tom\'{a}\v{s} Jungwirth}
\affiliation{School of Physics and Astronomy, University of Nottingham, University Park, Nottingham NG7 2RD, United Kingdom}
\affiliation{Institute of Physics, Academy of Sciences of the Czech Republic, Cukrovarnicka 10, 162 00 Praha 6, Czech Republic}
\author{Jonathan D. Denlinger}
\affiliation{Advanced Light Source, Lawrence Berkeley National Laboratory, Berkeley, California 94720, USA}
\author{Peter Wadley}
\affiliation{School of Physics and Astronomy, University of Nottingham, University Park, Nottingham NG7 2RD, United Kingdom}
\author{Dan Dessau}
\affiliation{Department of Physics, University of Colorado at Boulder, Boulder, CO 80309, USA}
\affiliation{Center for Experiments on Quantum Materials, Boulder, CO 80309, USA}
\date{\today}%

\maketitle

\section{\label{sec:DFT}DFT Calculations} 
The DFT calculations were performed with Vienna Ab-initio Simulation Package (VASP)~\cite{VASP}, based on a plane wave basis set. The interaction between ions and electrons is described by the projector-augmented wave (PAW) method. GGA+U scheme was applied, with the exchange-correlation functional being the Perdew-Burke-Ernzerhof generalized gradient approximation functional, and an on-site Coulomb potential of $2.25$ eV was applied to the Mn-3d electrons under the Dudarev approach. A full relaxation of the lattice structure was carried out with the energy and force breaking condition set to $10^{-6}$ eV and $10^{-3}$ eV/Å. The cut off energy was taken as $550$~eV. A dense k-mesh of 31*31*19 was used in the static calculation of the electronic structure. The calculations were carried out with each of the Mn atoms having a magnetic moment of 3.67~$\mu$B with the N\'eel vector along the [010] orientation. Finally, the chemical potential was shifted to bring the band top at $Z$ in the DFT to the corresponding maximum intensity in the experimental Energy Dispersion Cut (EDC) through $Z$. In the case of $U=2.25$~eV, this was $388$~meV, but the precise shift did depend on the specific value of $U$ used, with all U values requiring some shift.
\section{\label{sec:atomic_projected_bands} Atomically Projected Band Dispersion from DFT}
\begin{figure*}
    \includegraphics[]{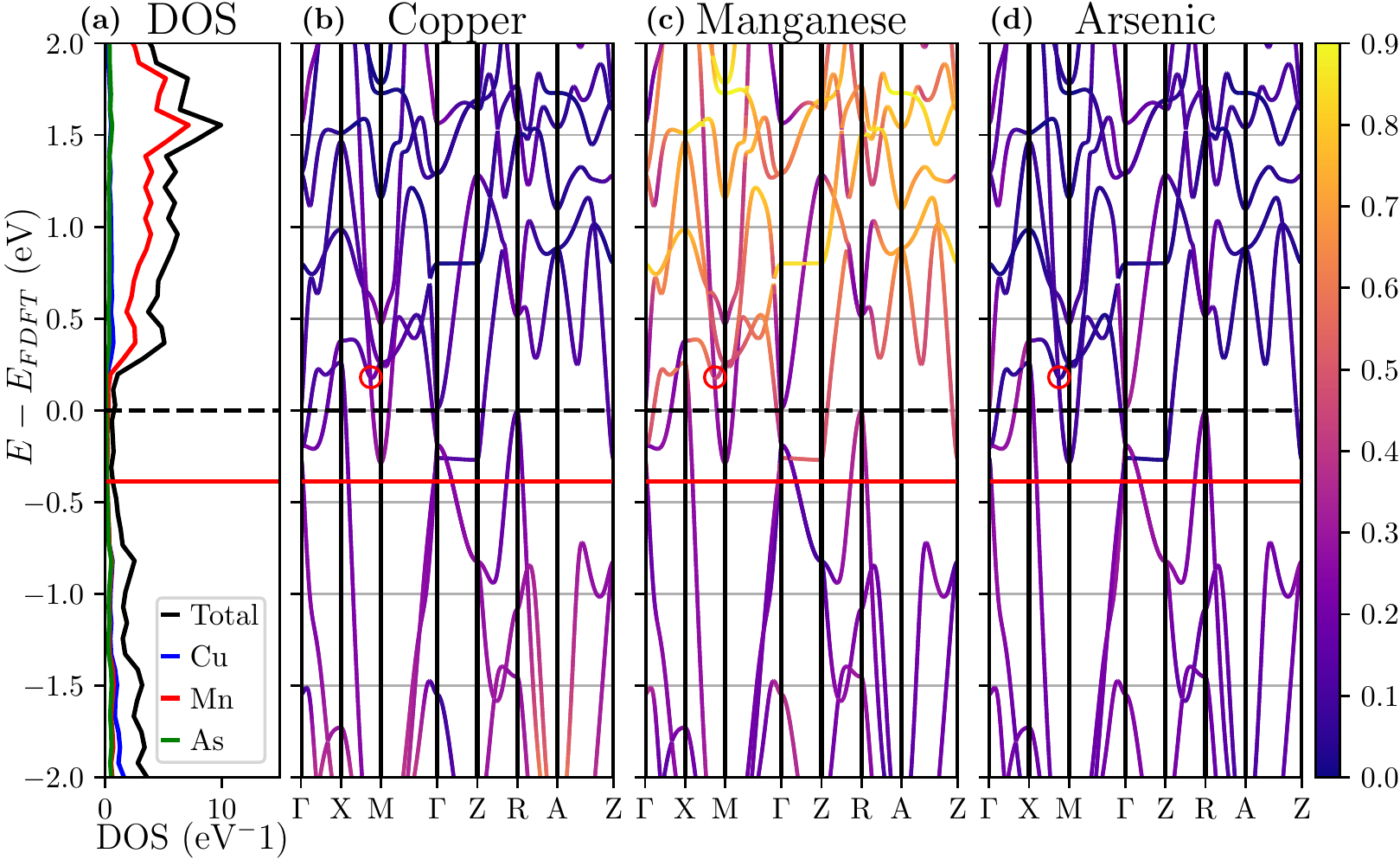}
    \caption{Atomic projected density of states and dispersion: (a) The Cu, Mn, As, and total DOS. (b,c,\&d) The electronic dispersion along high symmetry cuts with the color scale depicting the projection of the Bloch states onto Cu, Mn, and As atomic orbitals, respectively. In all figures, the black dashed lines indicate the Fermi energy from DFT, while the red lines show the experimental Fermi energy.}\label{Fig:atomic_projected_bands}
\end{figure*}

The atomic character of the bands predicted by the DFT are plotted with PyProcar~\cite{PyProcar} in Fig.~\ref{Fig:atomic_projected_bands}). There is strong hybridization of the bands near the experimental Fermi surface, $E-E_{FDFT}\approx-0.390$~eV on this unshifted energy scale (red line). This is in contrast to the pure GGA calculation where Mn dominated the Fermi surface. The application of $U=2.25$~eV to the Mn 3d orbitals has pushed the primary contribution from the Mn atoms above the Fermi surface and leveled the playing field between the three atomic species.
\section{\label{sec:FS_SOC} Fermi Surfaces with and without Spin-Orbit Coupling from DFT}
\begin{figure*}
    \includegraphics[]{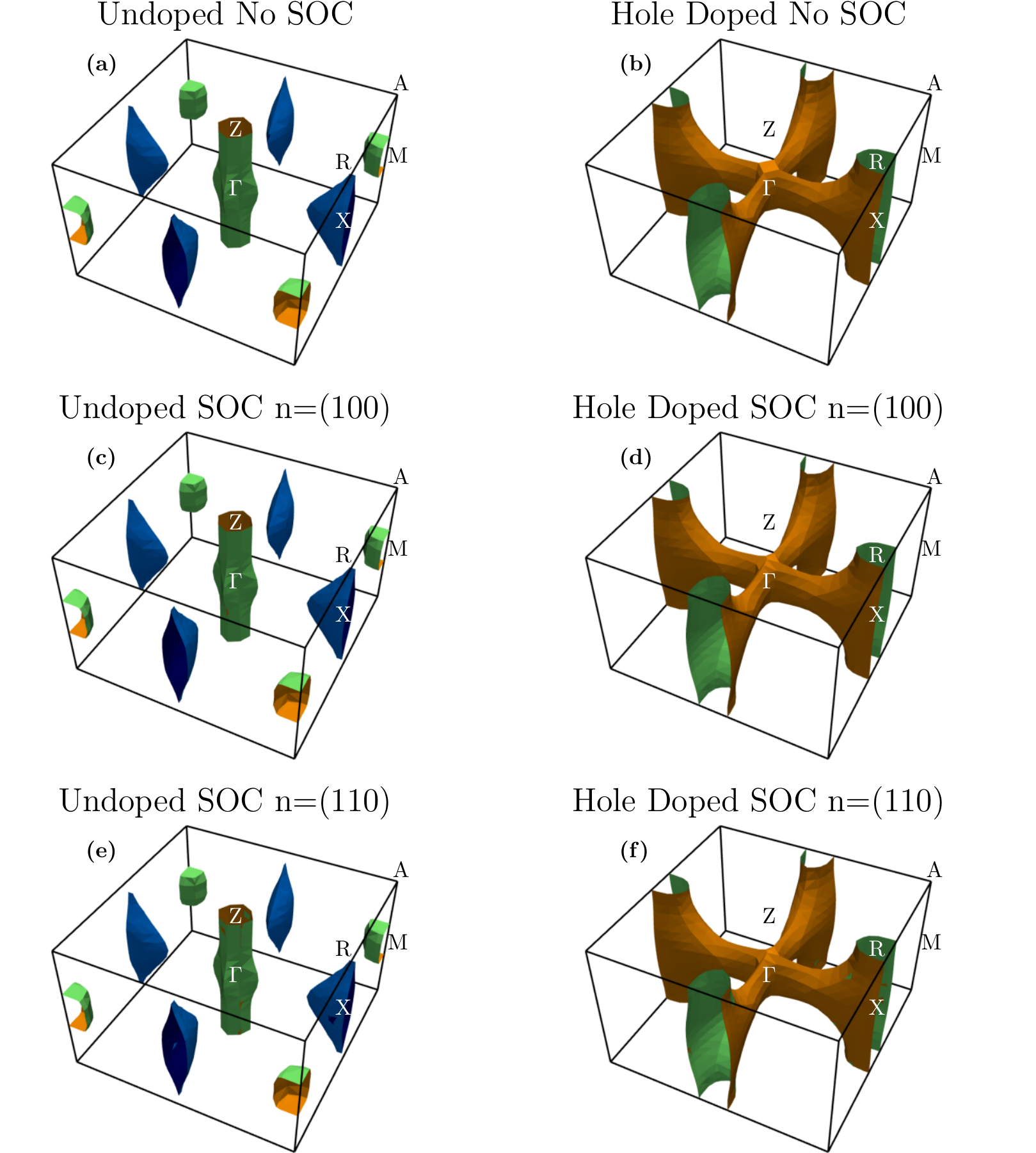}
    \caption{Fermi surfaces with and without SOC: Three dimensional Fermi surfaces are plotted at the stoichiometric Fermi energy calculated with DFT (a,c\&e) and at an energy $388$~meV below (hole doping) this energy (b,d,\&f), corresponding to our experimental $E_F$. The Fermi surfaces are from DFT calculations with no spin-orbit coupling (a\&b) and with spin-orbit coupling and the N\'eel vector fixed to the (1,0,0) direction (c\&d) and the (1,1,0) direction (e\&f). All calculations use GGA+U=2.25~eV, consistent with all other DFT shown, meaning that (a\&b) are the same as shown in Fig. 2 of the main text.}\label{Fig:FS_SOC}
\end{figure*}

To illustrate the effects of including spin-orbit coupling (SOC) on the Fermi surface, 3D Fermi surfaces are plotted with PyProcar~\cite{PyProcar} in Fig.~\ref{Fig:FS_SOC} with and without SOC included in the DFT calculations. The Fermi surfaces are plotted at the stoichiometric Fermi energy determined from DFT and at our experimental $E_F$, hole doped by $388$~meV relative to DFT. There is no visibly discernible distinction between the calculations with and without SOC, which is to be expected for the light magnetic Manganese atoms. As mentioned in the main text, it is necessary to account for SOC when calculating fermion masses, but because there are no Dirac crossing at the energies plotted, there is no appreciable difference in the Fermi surfaces from including SOC. 
\section{\label{sec:kz}$k_z$ Dispersion}
\begin{figure*}
    \includegraphics{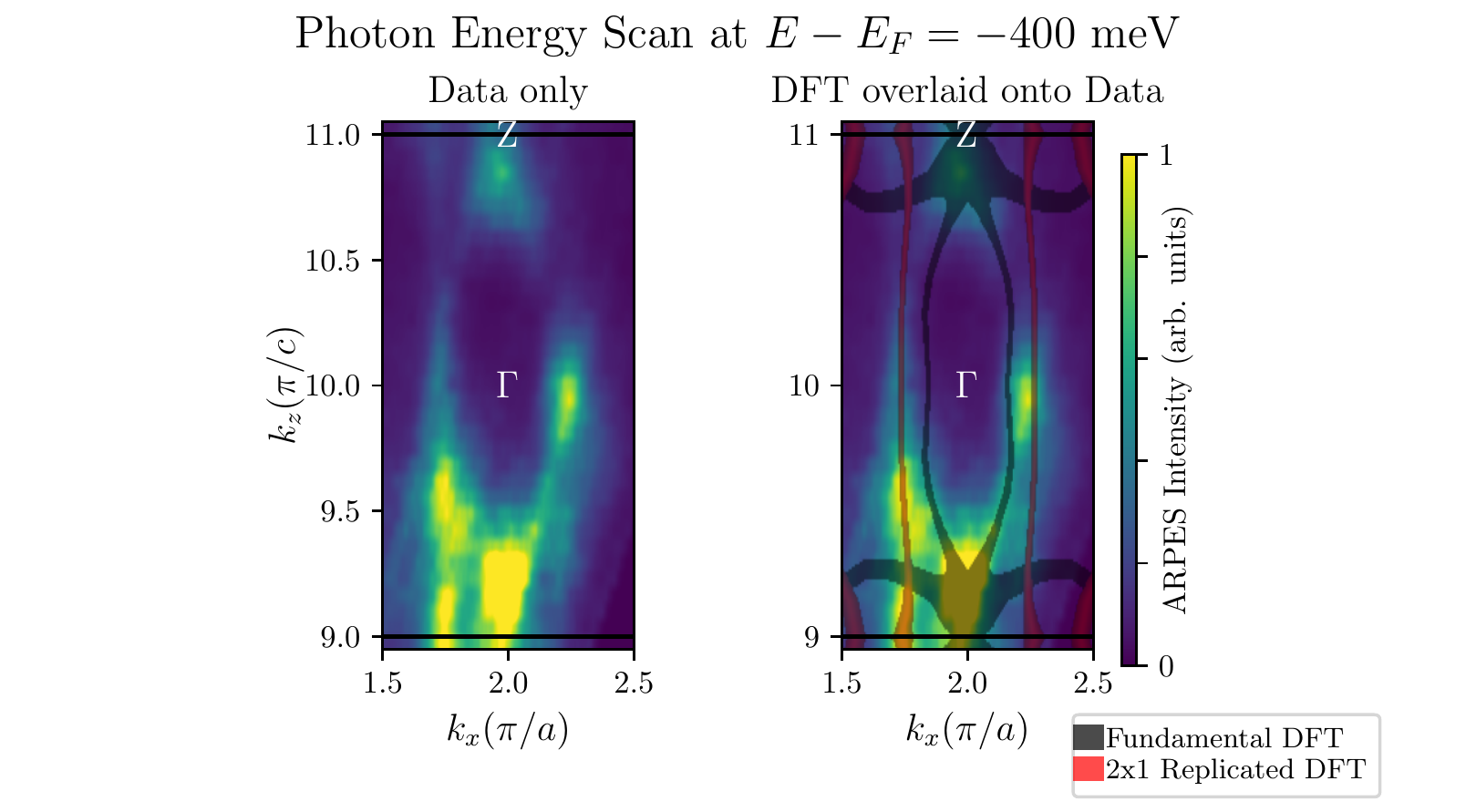}
    \caption{$k_z$ dispersion: (a\&b) Experimental data taken at ALS MERLIN ARPES. Using p-polarized light, photon energies were scanned between 70-125 eV while taking the $X\to \Gamma \to X$ cut; however, the data shown here is cropped to a single Brillouin zone (black box). Photon energies and angles are converted to $k_z$ and $k_x$ using an inner potential of 15~eV. High symmetry points are labeled in white text. (b) Same data as (a) but with the fundamental DFT overlaid in black transparency and the 2x1 replicated DFT overlaid with red transparency.}\label{Kz}
\end{figure*}
The $k_z$ dispersion was acquired by measuring along the $X \to \Gamma \to X$ high symmetry line while scanning the photon energy, $h\nu$, over the range $70$-$125$~eV in steps of $1$ eV, using p-polarized light. At these energies, the photon momentum may be neglected, yielding~\cite{Damascelli}
\begin{align}
k_z = \sqrt{\frac{2m_e}{\hbar}(h\nu-\Phi-E_b+V_0)-k_x^2-k_y^2}\label{kzEqn},
\end{align}
where $\Phi=4.5$ eV is the work function, $E_b$ is the binding energy, $k_x$ ($k_y$) is the in-plane momenta along the $a$ ($b$) direction of the lattice, and $V_0$ is the inner potential. The series of spectra over the photon energy range may be stitched together to yield the $k_z$-dispersion in the plane containing the $X \to \Gamma\to X$ cut (Fig.~\ref{Kz}). From Eqn.~\ref{kzEqn}, it can be seen that changing $V_{0}$ offsets the spectrum in $k_z$. Thus, the inner potential was tuned to produce the expected $k$-space periodicity in $k_z$, yielding $V_0 = 15$ eV. 

With the appropriate $k_z$ dispersion, the DFT can be overlaid onto the data as seen in Fig.~\ref{Kz}(b). The fundamental DFT captures the diamond-like structure enclosing $\Gamma$ quite nicely; however, there are additional features near $k_x\approx 1.8$ and $2.2 \pi/a$ that do not disperse in $k_z$. These 1D features are nicely explained by the same 2x1 replication seen in the LEED, Fermi surfaces, and the $R\to Z\to R$ symmetry cut. This is visually illustrated in Fig.~\ref{Kz}(b) by overlaying the replicated DFT in red. The non-dispersing feature can be identified, again, as the replication of the $k_y=\pi/a$ plane of k-space onto the $k_y=0$ plane. The circular pieces of the DFT are from the bands that make the tips of the ellipses in the $k_z\approx\pi/c$ Fermi surface, which for the polarization shown are suppressed by the matrix elements (see Sec.~\ref{sec:MatrixElementsFS}).

In addition to further testing the DFT against dispersion in $k_z$, determining the inner-potential, as done here, determines the $k_z$ value of any particular photon energy at $k_x$,$k_y$. In this way, the $k_z$ values of the data are labeled, using an ``$\approx$'' to acknowledge that the $k_z$ value is not constant across a cut nor Fermi surface and that the precise value from Eqn.~\ref{kzEqn} is not exactly $0$ or $\pi/c$. The $k_z$ values used to plot the DFT are labeled as approximate when the value from Eqn.~\ref{kzEqn} is used, yielding a slightly different value from the representative value listed. Furthermore, the exact distribution of $k_z$ values that contribute with varying strength to the spectral weight depends on the matrix elements and the details of the sample surface, yielding crisp dispersion in some zones for some bands and not for others. That is to say that determining the precise $k_z$ uncertainty for any particular cut is not practical. Hence a regular appeal to contributions from multiple $k_z$ values to explain spectral weight in various cuts and Fermi surfaces. 
\section{\label{sec:MatrixElementsFS}Matrix Elements in the Fermi Surface}
It is well known that ARPES spectral intensity is modulated by a transition matrix element that contains information about the orbital character of the band at a particular wave-vector $\boldsymbol{k}$. The norm squared of this matrix element in the dipole approximation can be shown to be proportional to
\begin{align}
    \bra{\phi^{\boldsymbol{k}}_{f}}\boldsymbol{\epsilon}\cdot\boldsymbol{r}\ket{\phi^{\boldsymbol{k}}_{i}}\label{eqn:M},
\end{align}
where $\boldsymbol{\epsilon}$ is the polarization vector of the incoming light, $\bra{\phi^{\boldsymbol{k}}_{f}}$ is the final state wave function of the photo-electron, and $\ket{\phi^{\boldsymbol{k}}_{i}}$ is the initial state wave function~\cite{Damascelli}. Importantly, the ARPES spectral intensity will go to zero if the implicit integrand in expression~\ref{eqn:M} is odd with respect to some mirror plane, because the odd integrand will integrate to zero.
\begin{figure*}
    \includegraphics{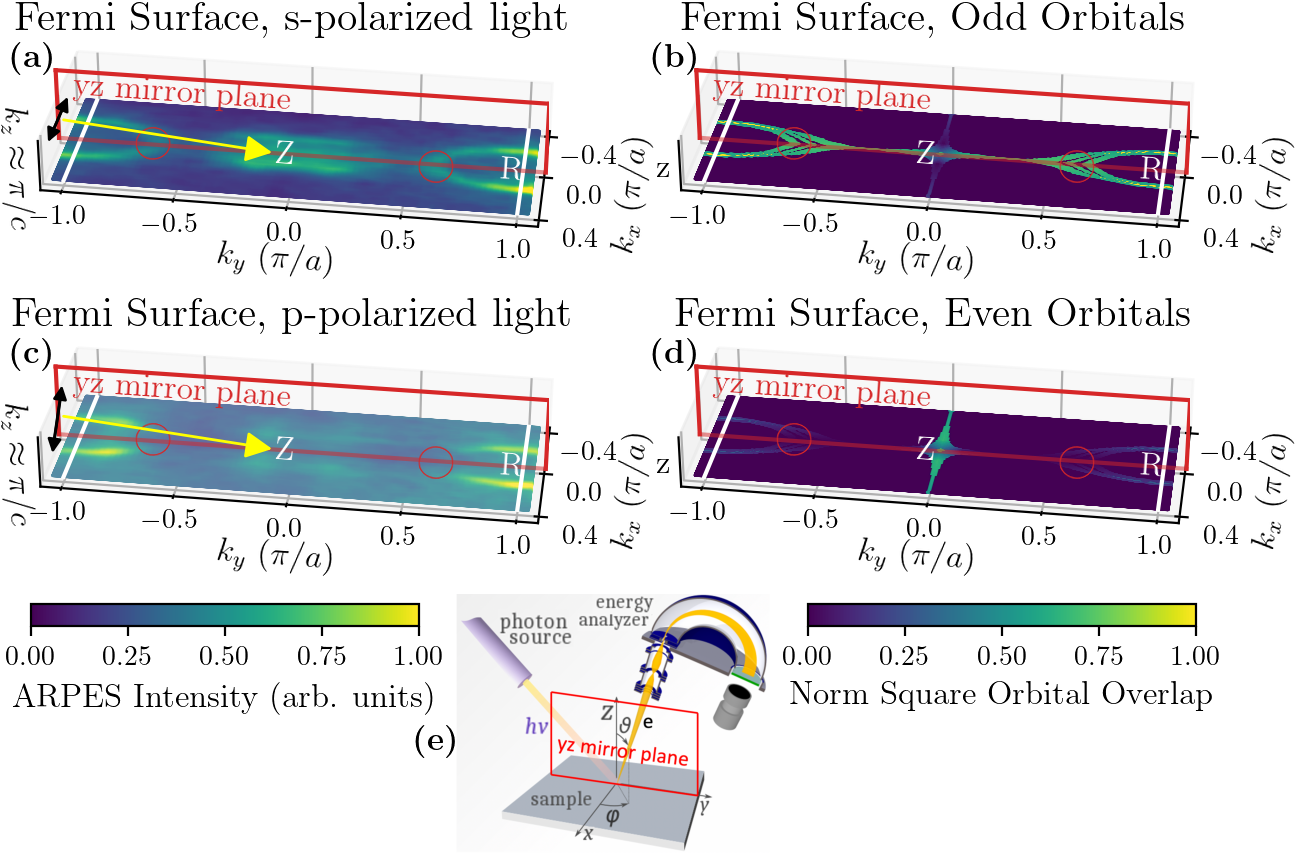}
    \caption{Orbital Decomposition: (a) Fermi surface data taken with 100 eV photons linearly polarized along $\boldsymbol{\hat{x}}$ [same data as Fig.~2(b) of the main text]. (c) Same as (a) but with the photon polarization lying in the yz mirror plane. (a\&c) Yellow arrow indicates incoming light, while black double arrows indicate the polarization of the light. (b\&d) Superimposed Fermi surface plots of the DFT at $k_z\approx\pi/c$, $\pi/c-0.16\pi/c$, and $\pi/c-0.28\pi/c$ projected onto the odd orbitals ($p_x$, $d_{xy}$, and $d_{xz}$) and even orbitals ($s$, $p_y$, $d_{yz}$, $d_{z^2}$, and $d_{x^2-y^2}$) with respect to the yz mirror plane, respectively. (a-d) White text labels symmetry points of the BZ. BZ boundaries indicated by white lines. (e) Experimental coordinate system~\cite{wiki:ExperimentalCoordinates}. (a-e) Red box indicates the yz mirror plane that contains incoming photon beam.} \label{OrbitalSymettries}
\end{figure*}

This effect can be seen in the experimental Fermi surface in the $k_z\approx\pi/c$ plane (see Fig.~\ref{OrbitalSymettries}). Taking the yz plane as the mirror plane, all final states emitted with $k_x=0$ are purely even with respect to the mirror plane (note that for any $k_x\neq0$ the final state is a mixture of odd and even). In the case of s-polarized light, $\boldsymbol{\epsilon}=\boldsymbol{\hat{x}}$, and thus $\boldsymbol{\epsilon}\cdot\boldsymbol{r}$ is odd. Therefore, only initial state wave functions with odd symmetry with respect to the yz mirror plane may be emitted. On the other hand, for p-polarized light, where $\boldsymbol{\epsilon}$ lies in the yz plane, $\boldsymbol{\epsilon}\cdot\boldsymbol{r}$ is purely even, and thus initial state wave functions that are odd will not be emitted in this mirror plane. With this in mind, the data clearly implies that the bands at $k_x\approx 0$ must be odd with respect to the yz plane, which includes the $p_x$, $d_{xy}$, and $d_{xz}$ orbitals. 

The DFT can be further scrutinized by comparing the orbital projection from the DFT to the polarization dependence of the experimental data. In Fig.~\ref{OrbitalSymettries}(b), the Fermi surface is projected onto the odd orbitals with respect to the yz plane from the DFT. In Fig.~\ref{OrbitalSymettries}(d), the same plot is shown but for the even orbitals: $s$, $p_y$, $p_z$, $d_{yz}$, $d_{z^2}$, and $d_{x^2-y^2}$. The DFT clearly shows the states at $k_x\approx 0$ are odd, with vanishing contributions from even states, which is consistent with the experimentally observed polarization dependence. Further examination of the DFT data indicates that all three odd states contribute comparably to the tips of the ellipses enclosed in the red circles. And for the states near the Z point along the red line, the $p_x$ and $d_{xz}$ states have comparable strength, while the $d_{xy}$ state's contribution is weak and vanishing at $Z$.

In Fig.~\ref{OrbitalSymettries}, the DFT is plotted for three $k_z$ values—$k_z\approx\pi/c$, $\pi/c-0.16\pi/c$, and $\pi/c-0.28\pi/c$—because the precise location of the tips of the ellipses and their sizes are dependent on the precise distribution of the $k_z$ values that contribute spectral weight. Additionally, the presence of the line through $Z$ is explained by the multiple $k_z$ values contributing. Fortunately, the parity of the orbital character of the bands is not as sensitive to $k_z$.
\section{\label{sec:SymmetryCuts} Symmetry Cuts}
\begin{figure*}
    \includegraphics{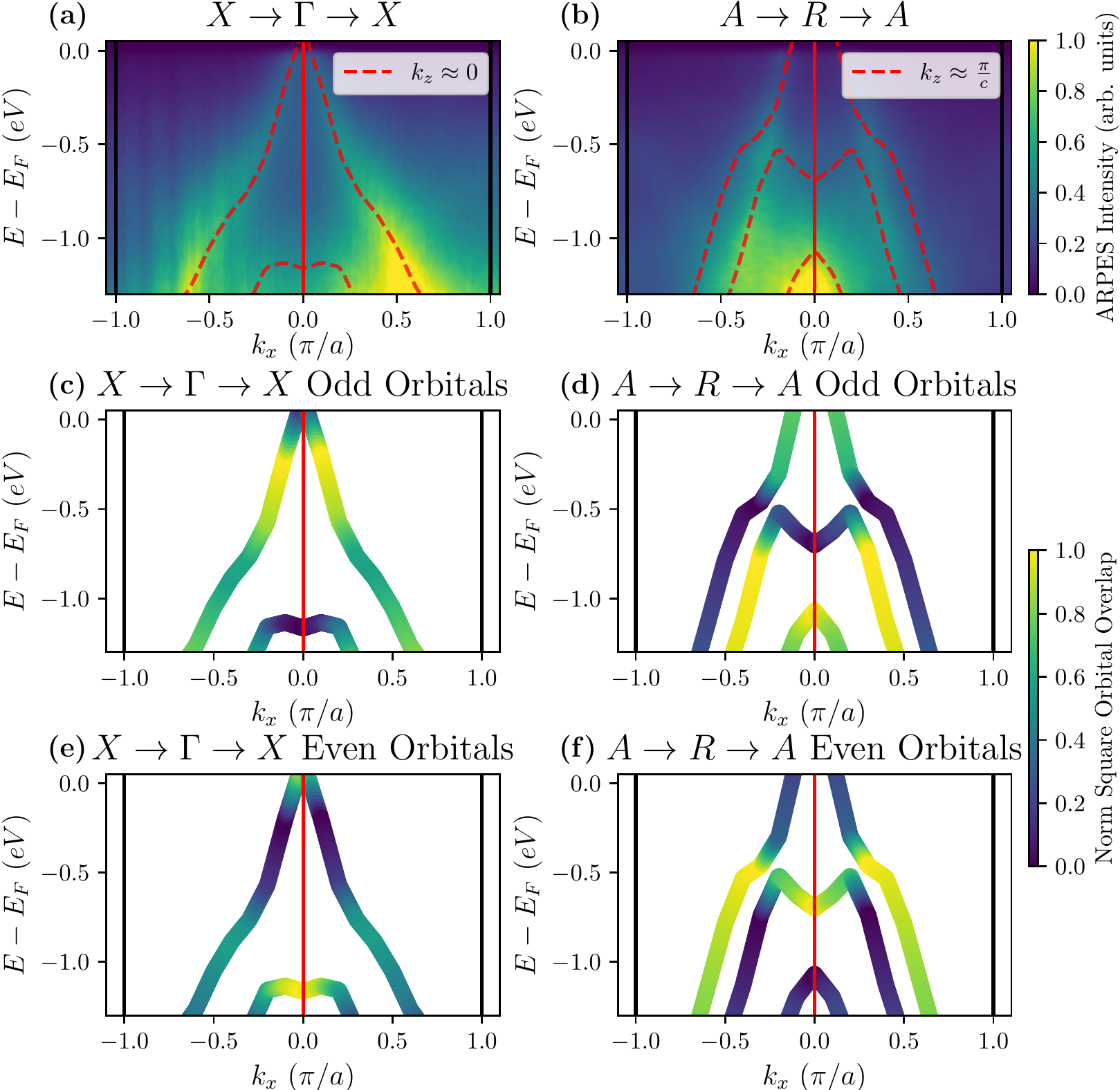}
    \caption{Matrix elements in symmetry cuts: (a) $X\to\Gamma\to X$ taken with 85 eV s-polarized ($\epsilon=\hat{x}$) photons. (b) $A\to R\to A$ symmetry cut was taken with 102 eV s-polarized photons. (a\&b) Red dashed lines show DFT. (c\&e) $X\to\Gamma\to X$ cut through the DFT projected onto odd (c) and even (e) orbitals with respect to the yz plane, which contains only $k_x=0$. (d\&f) Same as (c\&e) but for the $A\to R \to A$ cut through the DFT. (all) The solid red line indicates the states contained in the yz mirror-plane.}\label{Matrix_Elements}
\end{figure*}

In Fig.~\ref{Matrix_Elements}(a\&b), the symmetry cuts are seen to exhibit excellent qualitative agreement with the DFT. The band crossing $E_F$ in the $X\to\Gamma \to X$ cut tracks the DFT very well. The strong lines crossing $E_F$ in the $A\to R\to A$ cut are nicely tracked by the bands shown (the lack of a gap and subtle difference in $k_F$ is due to contributions from other $k_z$ values). Additionally, the band-top at $k_x=0$ in the $A\to R\to A$ cut shows good agreement.

There are, however, bands in the $X\to\Gamma\to X$ and $A\to R\to A$ cuts from the DFT that do not appear in the data, which under careful inspection reveal a deeper agreement between theory and experiment on the orbital content of the missing bands. Figure~\ref{Matrix_Elements} demonstrates the matrix element effects that are suppressing these bands. 

In both cases, the relevant mirror plane to consider is the yz plane, which contains the final states (even) for \emph{only} $k_x=0$. The data was taken with polarization along $\hat{x}$ (odd); so the matrix element sends the spectral intensity to zero for even initial states with respect to the yz mirror plane. The portion of the band-top near $k_x\approx 0$ and $E-E_F\approx -1.2$~eV in the $X\to \Gamma\to X$ cut is suppressed in the data. Additionally, the band bottom at $k_x=0$ in the $A\to R\to A$ cut is suppressed. Both are consistent with the prediction from the DFT that these bands at $k_x=0$ consist entirely of even orbitals.
\section{\label{sec:uncertainty}Fermi Velocity Uncertainty from MDC Fitting}
\subsection{General Intuition}
In MDC fitting, the energies, $\set{E}$, of the MDCs are assumed to be precisely the measured values. However, there will be uncertainty in the extracted peak positions, $\set{k}$, related to instrument resolution and uncertainty in the fitting due to noise in the data. This uncertainty will lead to an uncertainty in the Fermi velocity. The scaling of the uncertainty in $v_F$ can be illustrated by considering
\begin{align}
    \begin{split}
        v_F &= \frac{dE}{dk}\Big|_{k_{F}} \\
        &\approx \frac{\Delta E}{\Delta k \pm\sigma_{\Delta k}}\bigg|_{k_{F}} \\
        &= \frac{\Delta E}{\Delta k}\bigg|_{k_{F}} \pm\sigma_{v_F}
    \end{split}
\end{align}
where $\Delta k$ $(\Delta E)$ is just the difference between the two $k$ ($E$) points closest to the Fermi energy and $\sigma_{\Delta k}$ is the uncertainty in the difference of the two k-points. For small $\sigma_{\Delta k}$, the resulting uncertainty in $v_{F}$, $\sigma_{v_F}$, can be found by transforming the variance in $\Delta k$ under the Jacobian:
\begin{align}
    \sigma_{v_F}^{2} &\approx \frac{dv_{F}}{d\Delta k} \sigma_{\Delta k}^{2} \frac{dv_{F}}{d\Delta k}\\
    \begin{split}
        \sigma_{v_F} &=|\Delta E (\Delta k)^{-2}\sigma_{\Delta k}| \\
        &= |v_{F}^{2}\frac{\sigma_{\Delta k}}{\Delta E}|.
    \end{split}
\end{align}
So, the uncertainty in $v_F$ scales as $v_{F}^{2}$. 
\subsection{Particulars of Our Method}
First, the $\set{E}$, $\set{k}$, and $\set{\sigma_{k}}$ that correspond to a band must be extracted. Since the bands within 150 meV of $E_F$ are highly linear, the ARPES spectral intensity from a band along an MDC can be approximated as a Voigt function in k. So, MDCs for all energies, $\set{E}$, between $E_F$ and $E_F-150$~meV are fit using a linear combination of Voigt functions to represent the bands. The center locations of the Voigt functions are the $\set{k}$. Although the $\set{\sigma_{k}}$ can be acquired from the non-linear least squares fitting directly, a better estimate for $\set{\sigma_{k}}$ can be found by using a Markov chain Monte Carlo simulation. This algorithm explores the parameter space nearby the best fit solutions to calculate the posterior distribution of the model parameters. Using the calculated probability distributions, $\set{\sigma_{k}}$ can be estimated as half the distance between the 15.8 and 84.2 percentiles of the marginal probability distribution of the peak location, $k$. All of this fitting is done in Python using Lmfit~\cite{LMFit}. 

Second, the uncertainty in $v_{F}$ from the derivative method given above can be improved by fitting polynomials to the pairs of $\set{E}$ and $\set{k}$ corresponding to a band. In particular, linear polynomials are fit to extract $k(E) = b*E + c$ where $k$ is treated as the dependent variable for technical reasons, including that $E$ varies more than $k$ for large Fermi velocities. $k(E)$ is fit with standard least squares fitting, using Numpy's Polyfit function~\cite{Numpy}, which calculates the covariance matrix over $b$ and $c$ taking into account $\set{\sigma_{k}}$. 

Finally, $v_{F} = 1/b$. Using $\sigma_b$ from the covariance matrix of $b$ and $c$, the uncertainty in $v_{F}$ can be found by again transforming $\sigma_{b}$ under the Jacobian: 
\begin{align}
    \sigma_{v_F}^{2} &\approx \frac{dv_{F}}{db} \sigma_{b}^{2} \frac{dv_{F}}{db}\\
    \begin{split}
        \sigma_{v_F} &=|\frac{1}{b^2}\sigma_{b}| \\
        &= |v_{F}^{2}\sigma_{b}|.
    \end{split}
\end{align}
Again, $\sigma_{v_{F}}\sim v_{F}^{2}$.

In this way, the Fermi velocities of the bands in the experimental ARPES cuts are determined: $X\to\Gamma\to X$ has $v_F = 4.6 \pm 0.5$eV\si{\angstrom}, $M\to X\to M$ has $v_F = 6.0 \pm 0.7$eV\si{\angstrom}, and $R\to Z\to R$ has $v_F = 5.1 \pm 0.5$eV\si{\angstrom}. Hence, the reported relative uncertainty in the main text of $\sim10\%$. Note, that the uncertainty in the $M\to X\to M$ cut cannot be reliably found from the Markov chain Monte Carlo simulations due to the lack of convergence in a reasonable amount of time. This is caused by the presence of four partially overlapping peaks. Therefore, the uncertainty found in the same manner outlined above, except using $\set{\sigma_k}$ acquired directly from the non-linear least squares fitting of the MDCs, is reported.

\bibliographystyle{unsrt}
\bibliography{./CuMnAs.bib}